\documentstyle[aps,psfig,prl]{revtex}
\begin{document}
\draft
\twocolumn[\hsize\textwidth\columnwidth\hsize\csname
@twocolumnfalse\endcsname
\title{CAPITAL REDISTRIBUTION BRINGS WEALTH BY PARRONDO'S PARADOX}

\author{RA\'UL TORAL}
\address{Instituto Mediterr\'aneo de Estudios Avanzados, IMEDEA (CSIC-UIB)\\Campus UIB, 07071-Palma de Mallorca, Spain\\
\centerline{\footnotesize\it email: raul@imedea.uib.es,  http://www.imedea.uib.es/PhysDept}} 
\date{\today}
\maketitle

\begin{abstract}We present new versions of the Parrondo's paradox by which a losing game can be turned into winning by including a mechanism that allows redistribution of the capital amongst an ensemble of players. This shows that, for this particular class of games, redistribution of the capital is beneficial for everybody. The same conclusion arises when the redistribution goes from the richer players to the poorer.\end{abstract}
]

Parrondo's paradox\cite{ha99,ha99b,hatp00,hatpp00} shows that the combination of two losing games does not necessarily generate losses but can actually result on a winning game. The paradox translates into the language of very simple gambling games (tossing coins) the so--called {\sl ratchet effect}, namely, that it is possible to use random fluctuations (noise) in order to generate {\sl ordered} motion against a potential barrier in a nonequilibrium situation\cite{ab94}. In this paper we introduce a new scenario for the Parrondo's paradox which involves a set of players and where one of the games has been replaced by a redistribution of the capital owned by the players. It will be shown that even though each individual player (when playing alone) has a negative winning expectancy, the redistribution of money brings each player a positive expected gain. This result holds even in the case that the redistribution of capital is directed from the richer to the poorer, although in this case the distribution of money amongst the players is more uniform and the total gain is less. 

Our games will consider a set of $N$ players. They are randomly chosen for playing. In player $i$'s turn ($i=1,\dots,N$) at time $t$, a (probably biased) coin is tossed such that the player's capital $C_i(t)$ increases (decreases) by one unit if heads (tails) show up. The total capital is $C(t)=\sum_i C_i(t)$. Time $t$ is measured in units of tossed coins per player and games are classified as winning, losing or fair if the average capital $\langle C(t)\rangle$ increases, decreases or remains constant with time, respectively. 

Let us start by reviewing briefly two versions of Parrondo paradox. Both of them consider a single player, $N=1$, but differ in the rules of one of the games:

{\bf Version I}: This is the original version\cite{ha99}. It uses two games, $A$ and $B$. For \underline{\bf game $A$} a single coin is used and there is a probability $p$ for heads. Obviously, game A is fair if $p=1/2$. \underline{\bf Game $B$} uses two coins according to the current value of the capital: if the capital $C(t)$ is a multiple of $3$, the probability of winning is $p_1$, otherwise, the probability of winning is $p_2$. The condition for $B$ being a fair game turns out to be $(1-p_1)(1-p_2)^2=p_1p_2^2$. Therefore, the set of values $p=0.5-\epsilon$, $p_1=0.1-\epsilon$, $p_2=0.75-\epsilon$, for $\epsilon$ a small positive number, is such that both game A and game B are losing games. However, and this is the paradox, a winning game is obtained for the same set of probabilities if games A and B are played randomly by choosing with probability $1/2$ the next game to be played\cite{alternation}. 

{\bf Version II}: This version of the paradox \cite{pha00} eliminates the need
for using modulo rules based on the player's capital, which are of difficult
practical application. It keeps game A as before, but it  modifies game B to a
new \underline{\bf game B'} by using four different coins (whose heads
probabilities are $p_1$, $p_2$, $p_3$ and $p_4$) at time $t$ according to the
following rules: use (a) coin $1$ if game at $t-2$ was loser and game at $t-1$
was loser; (b) coin $2$, if game at $t-2$ was loser and game at $t-1$ was
winner; (c) coin $3$, if game at $t-2$ was winner and game at $t-1$ was loser;
(d) coin $4$ if game at $t-2$ was winner and game at $t-1$ was winner. The
condition for the game B' to be a fair one is $p_1p_2=(1-p_3)(1-p_4)$. The
paradox appears, for instance, choosing $p=1/2-\epsilon, p_1=0.9-\epsilon$,
$p_2=p_3=0.25-\epsilon$, $p_4=0.7-\epsilon$, for small positive $\epsilon$,
since it results in A and B' being both losing games but the random alternation
of A and B' producing a winning result.

This type of paradoxical results has been found in other cases, including work
on quantum games\cite{na02}, pattern formation\cite{blp02}, spin
systems\cite{m00}, lattice gas automata\cite{mb02}, chaotic dynamical
systems\cite{affm02}, noise induced synchronization\cite{tmhp01,kt02},
cooperative games\cite{t01},  and possible implications of the paradox in other
fields, such as Biology, Economy and Physics\cite{d01}.

In this work we consider an ensemble of players and replace the randomizing
effect of game A by a redistribution of capital amongst the players. In
particular, we have considered $N$ players playing versions I and II as
modified by the following rules:

{\bf Version I'}: A player $i$ is selected at random for playing. With probability $1/2$ he can either play game B or \underline{\bf game A'} consisting in that player giving away one unit of his capital to a randomly selected player $j$. Notice that this new game A' is fair since it does not modify the total amount of capital, it simply redistributes it randomly amongst the players. 

{\bf Version II'}: It is the same than version I' but with the modulo dependent game B replaced by the history dependent game B'. 
\begin{figure}
\centerline{\psfig{file=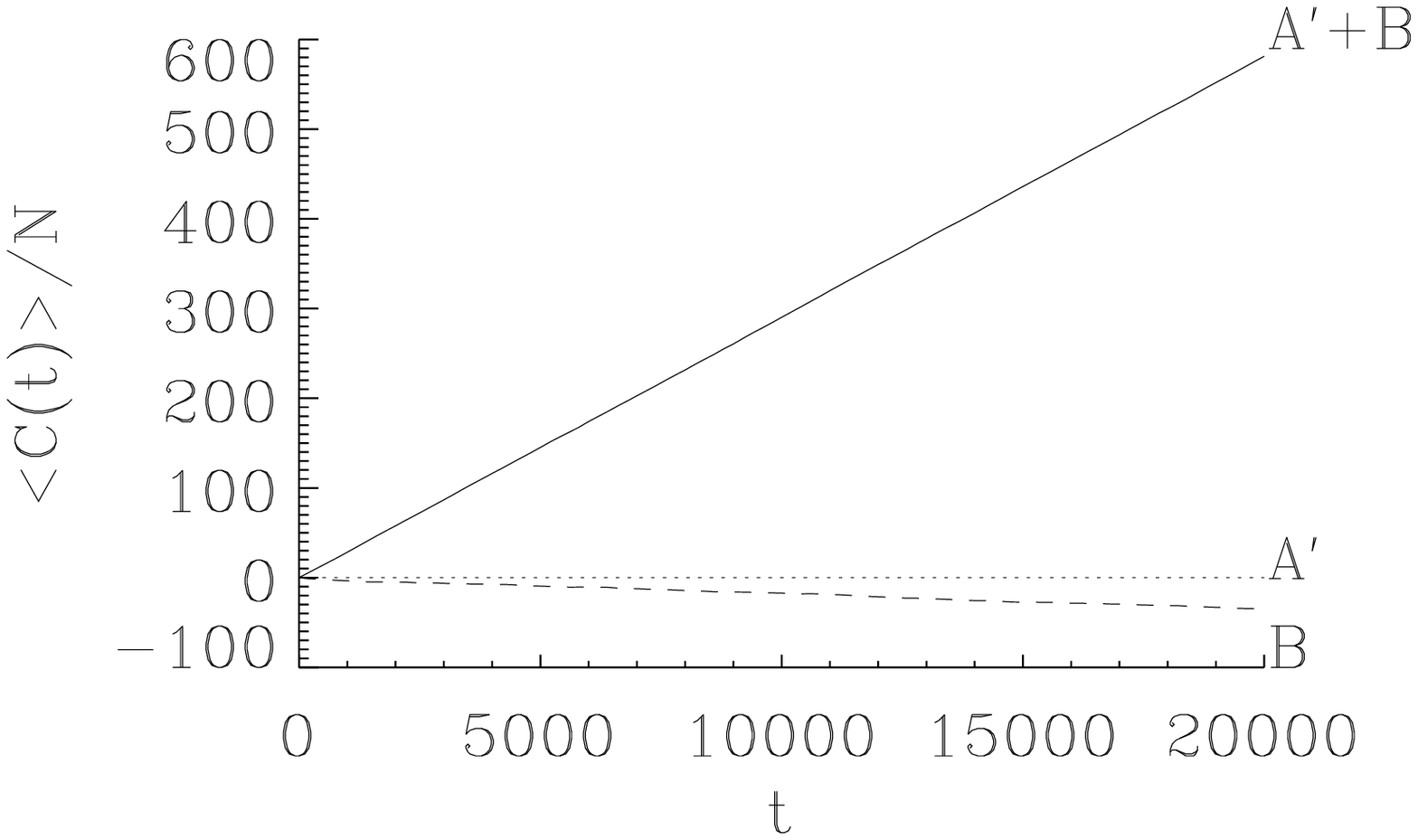,width=9.0cm,height=6.0cm}} 
\vspace{-25.0pt}
\centerline{\psfig{file=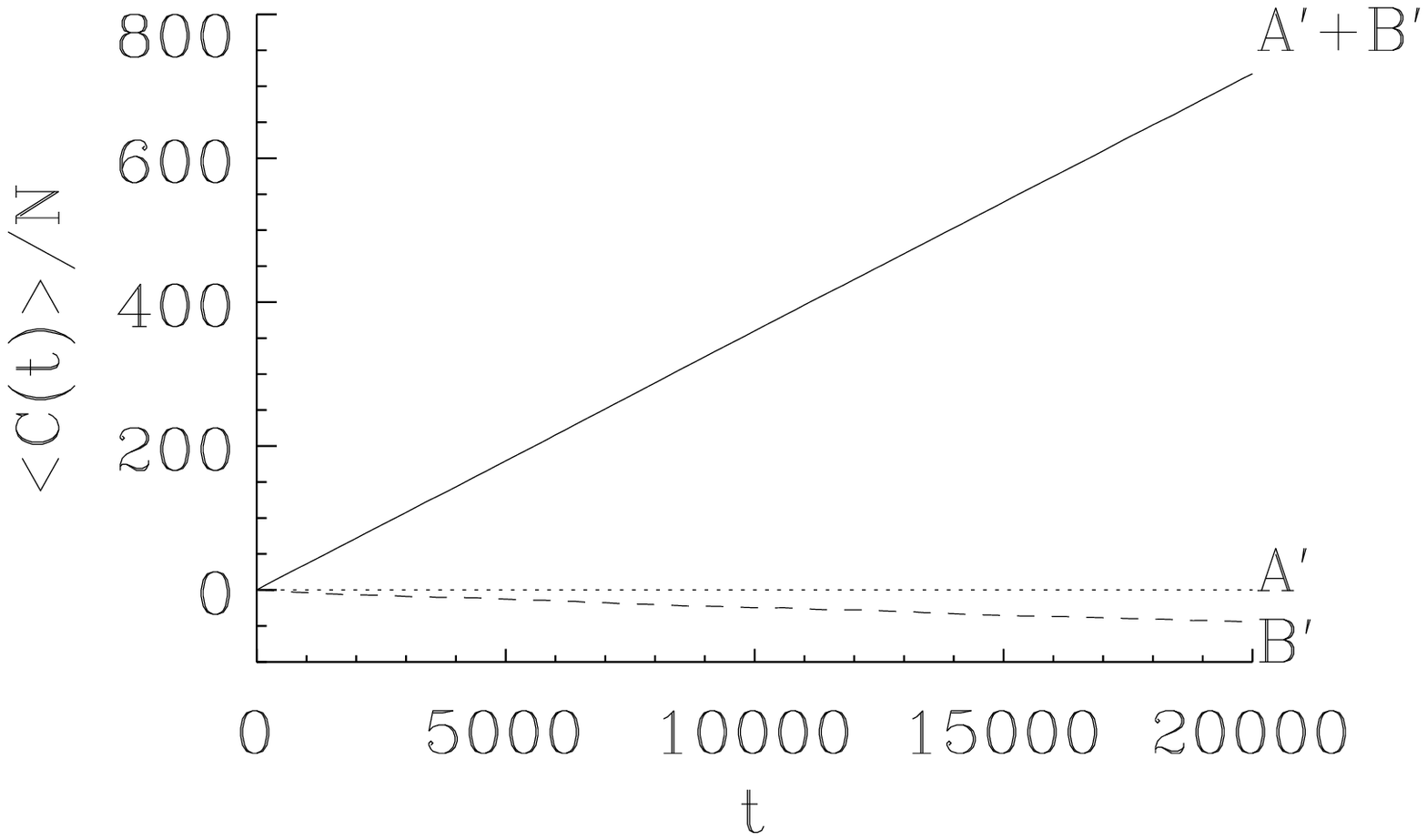,width=9.0cm,height=6.0cm}} 
\vspace{-25.0pt}
\centerline{\psfig{file=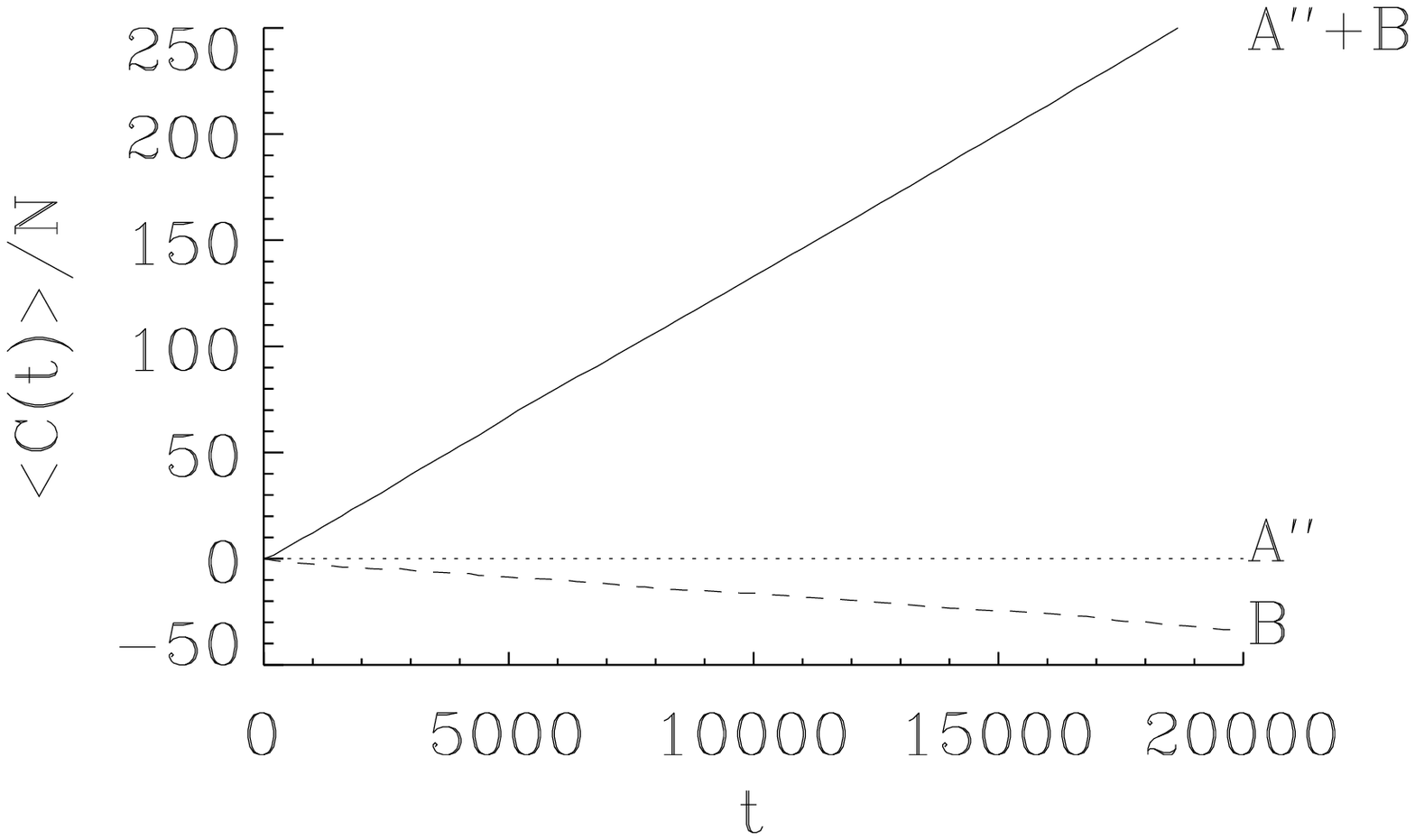,width=9.0cm,height=6.0cm}} 
\caption{Average capital per player, $\langle C(t)\rangle/N$, versus time, $t$, measured in units of games per player. The different games A', A'', B and B' are described in the main text. The probabilities defining the games are as follows: $p_1=0.1-\epsilon,\,p_2=0.75-\epsilon$ for game B; $p_1=0.9-\epsilon,\, p_2=p_3=0.25-\epsilon,\, p_4=0.7-\epsilon$ for game B', with $\epsilon=0.01$ in both games. We consider an ensemble of $N=200$ players and the results have been averaged for $10$ realizations of the games. In all cases, the initial condition is that of zero capital, $C_i(0)=0$, for all players, $i=1,\dots,N$. Notice that while games A' and A'' are fair (zero average) and games B and B' are losing games, the random alternation between games as indicated by A'+B (top panel), A'+B'(middle panel) and A''+B (bottom panel) result in winning games.
\label{fig1}} 
\end{figure}

\begin{figure}
\centerline{\psfig{file=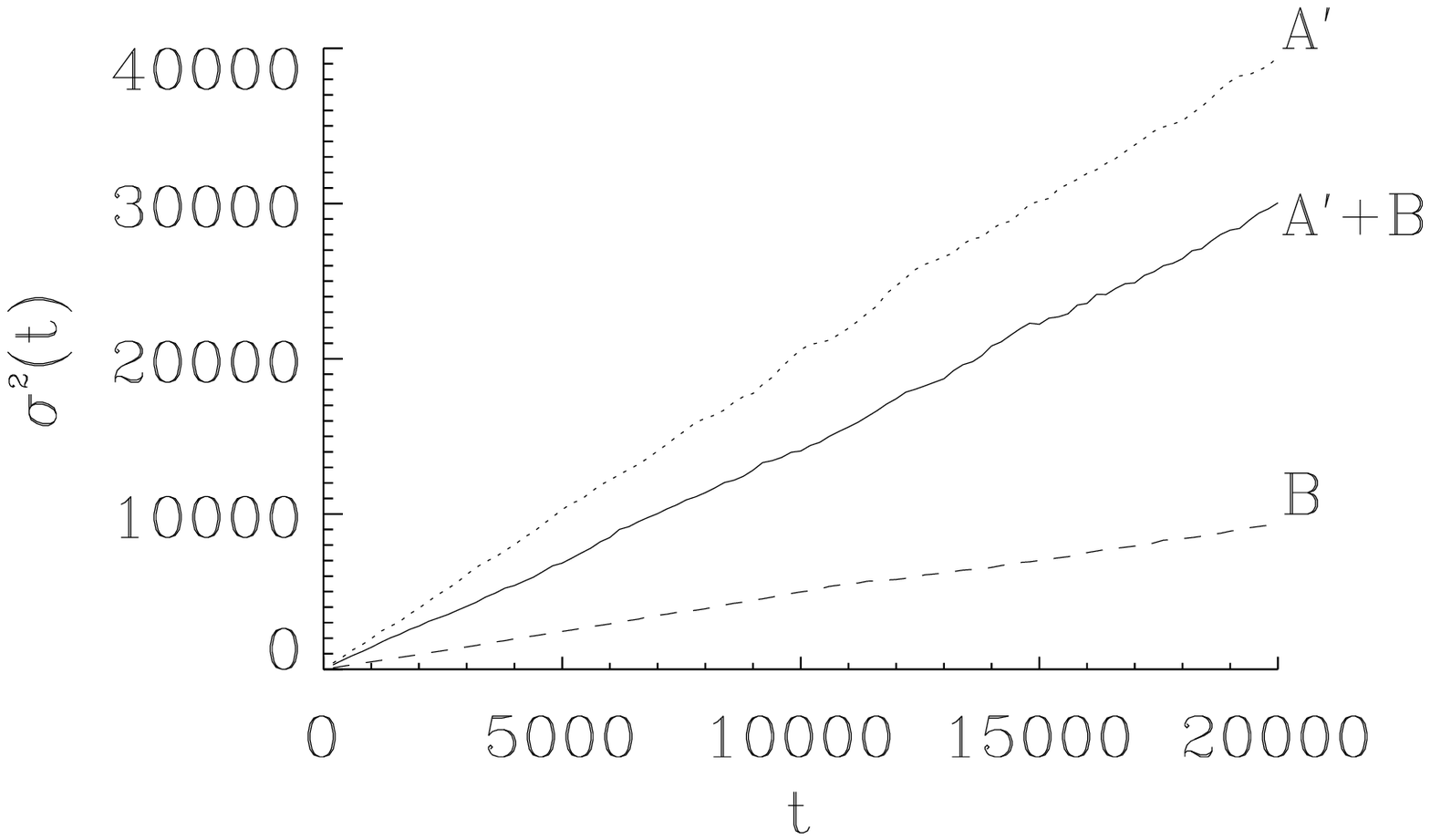,width=9.0cm,height=6.0cm}}
\vspace{-20.0pt}
\centerline{\psfig{file=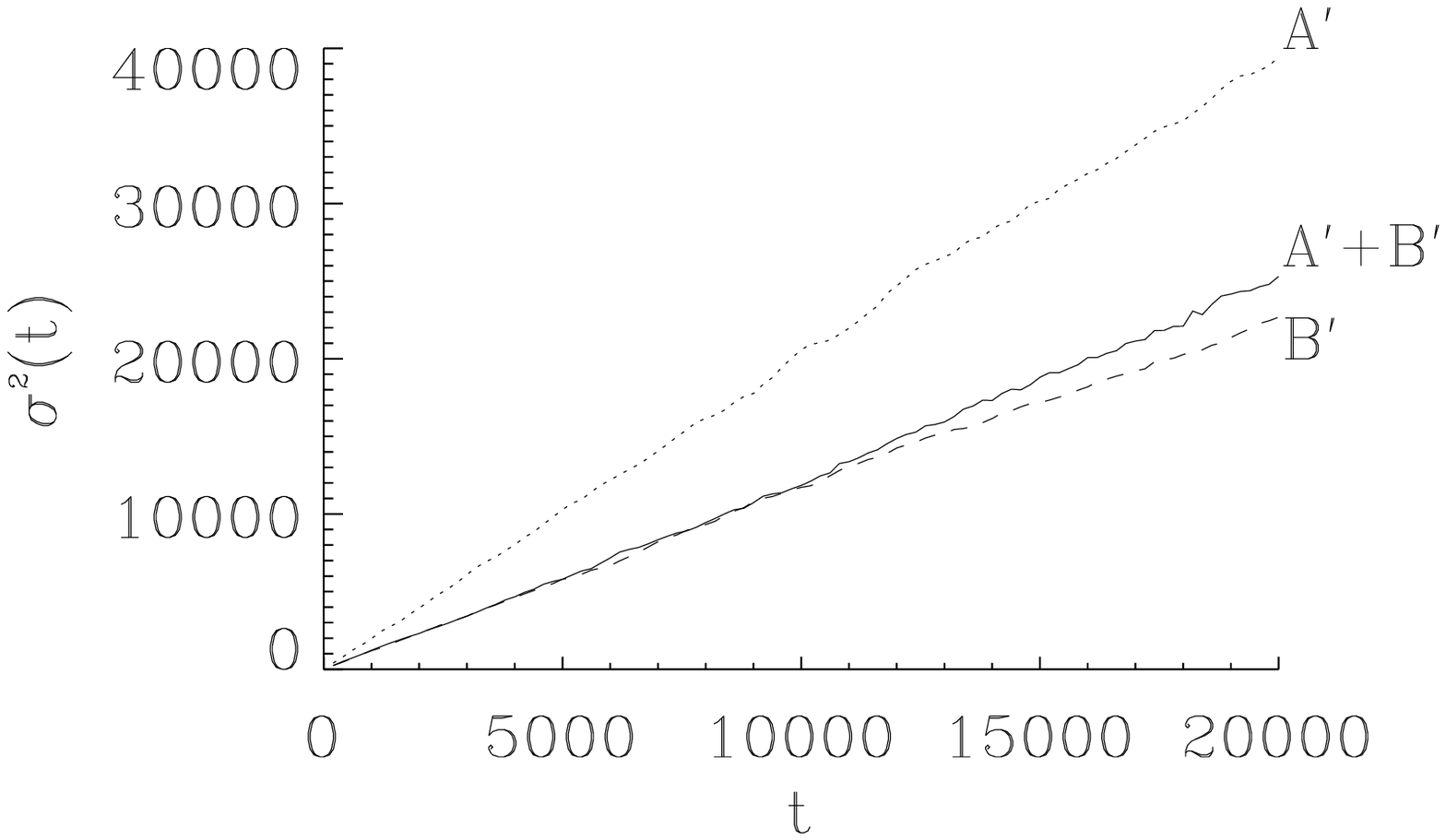,width=9.0cm,height=6.0cm}}
\vspace{-20.0pt}
\centerline{\psfig{file=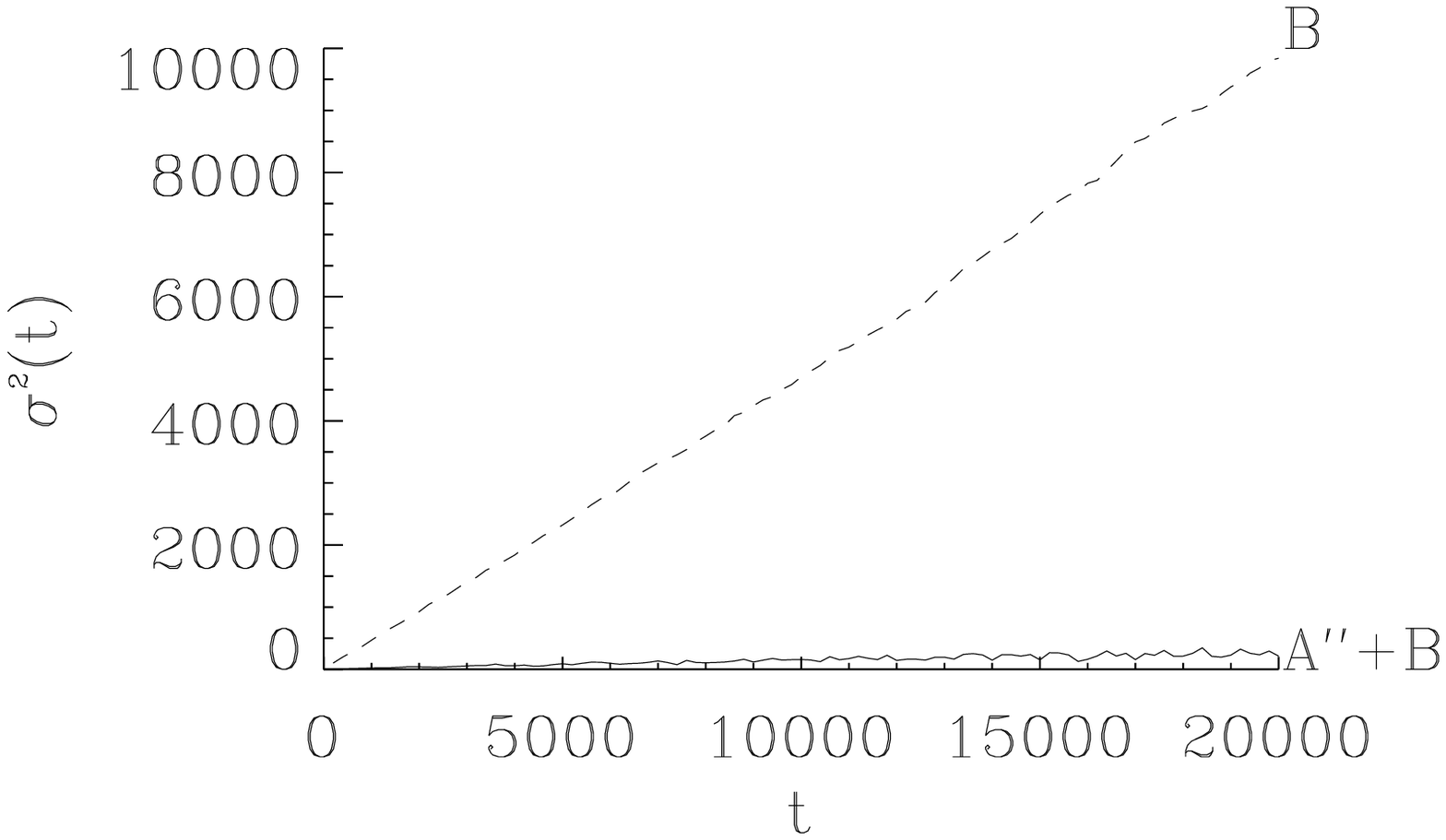,width=9.0cm,height=6.0cm}}
\caption{Time evolution of the variance $\sigma^2(t)=\frac{1}{N}\sum_iC_i(t)^2-\left(\frac{1}{N}\sum_iC_i(t)\right)^2$ of the single player capital distribution in the same cases than in figure 1. 
\label{fig2}}
\end{figure}

\begin{figure}
\centerline{\psfig{file=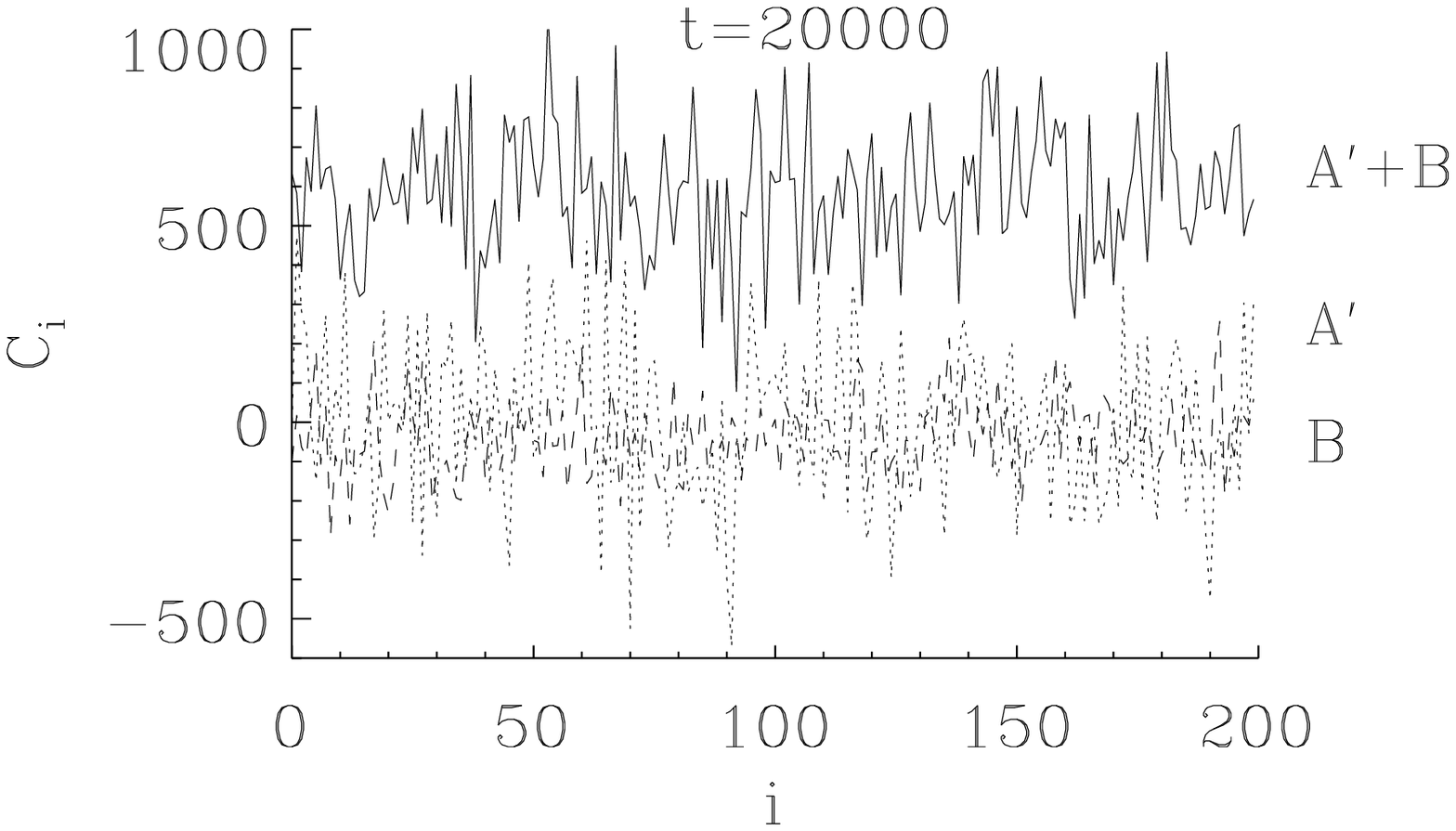,width=9.0cm,height=6.0cm}}
\vspace{-20.0pt}
\centerline{\psfig{file=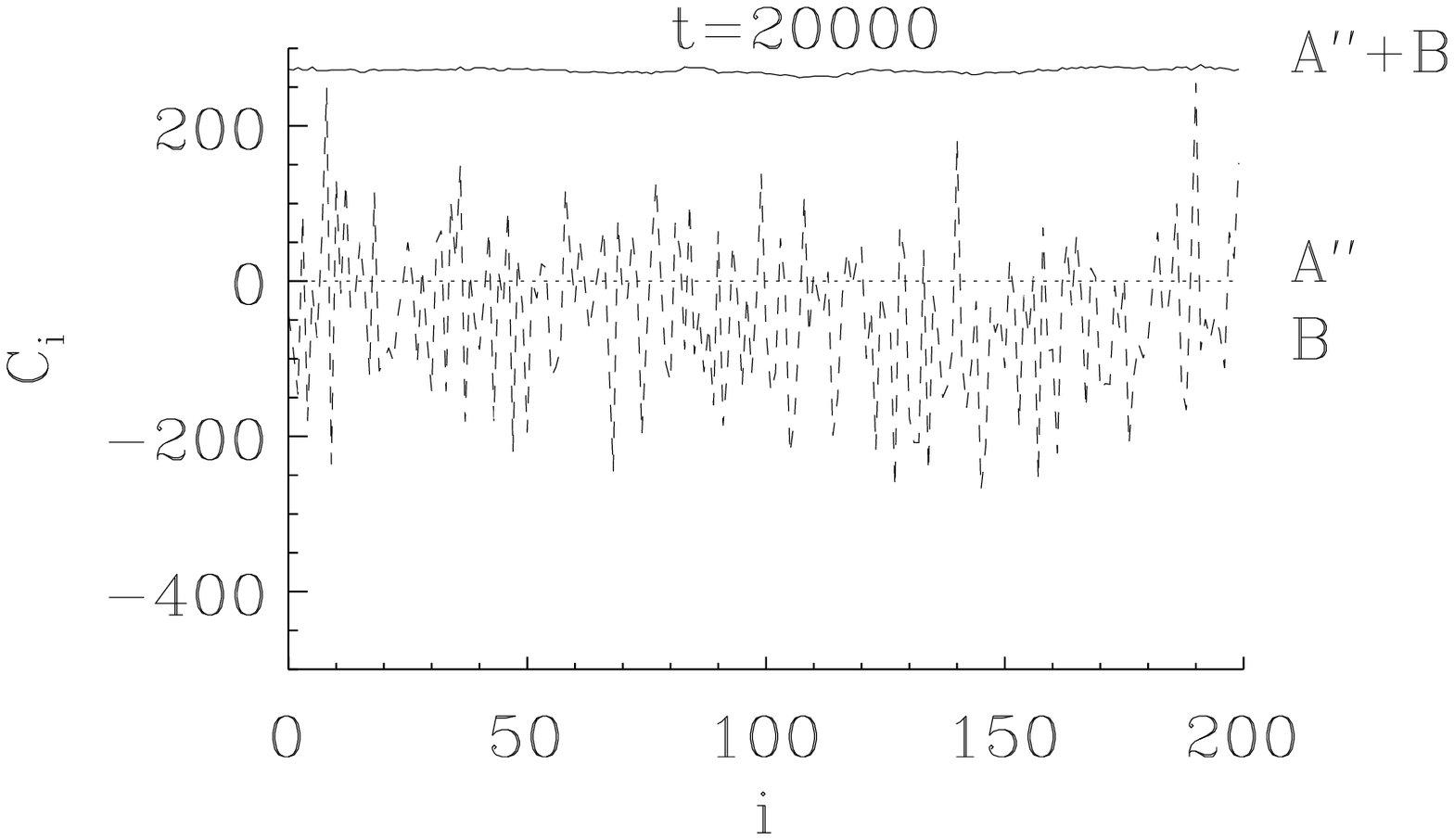,width=9.0cm,height=6.0cm}}
\caption{Capital distribution for an ensemble of $N=200$ players after a time $t=20000$ in the cases of combination of games A' and B (top) and games A'' and B (bottom) (same line meanings that in previous figures). Notice the almost flat distribution of money in the latter case. 
\label{fig3}}
\vspace{20.0pt}
\end{figure}

As it is shown in figure (\ref{fig1}), the Parrondo paradox appears for both versions I' and II'. It is clear from these figures that the random alternation of games A' and B or games A' and B' produces a winning result, whereas any of the games  B and B', played by themselves are losing games and game A' is a fair game. This proves that the redistribution of capital can turn a losing game into a winning one. In other words, it turns out to be more convenient for players to give away some of their money at random instants of time. This surprising result shows that a mechanism of redistribution of capital can actually, and under the rules implied in the simple games analyzed here, increase the amount of money of all the ensemble. This can be more shocking when we realize that the redistribution can be made from the richer to the poorer players, while still obtaining the paradoxical result. To prove this, we have replaced game A' by yet another \underline{\bf game A''} in which player $i$ gives away one unit of its capital to any of its nearest neighbors with a probability proportional to the capital difference. To be more precise, the probability of giving one unit from player $i$ to player $i+1$ or to player $i-1$ is $P(i\to i\pm 1)\propto \max[C_{i}-C_{i\pm 1},0]$, with $ P(i\to i+1)+P(i\to i-1)=1$. These probabilities implies that capital always goes from one player to a neighbour one with a smaller capital and never otherwise. These rules are in some sense, similar to the ones used in solid on solid type models to study surface roughening\cite{ks92}. Under the only influence of game A'', the capital is conserved and tends to be uniformly distributed amongst all the players.

We now study the variance of the capital distribution amongst the players. The results, plotted in figure (\ref{fig2}), show that the variance of the capital distribution of the random combinations of game A' with games B or B' lies always in between of the individual games. This proves that the overall increase of capital observed in the random combination of games is not obtained as a consequence of a very irregular distribution of the capital amongst the players. In the combination A''+B the homogenization effect of game A'' brings a nearly uniform distribution of capital amongst the players, see figure (\ref{fig3}).

In conclusion, we have introduced new versions of the Parrondo's paradox which involve an ensemble of players and rules that allow the redistribution of capital amongst the players. It is found that this redistribution (which by itself, has no effect in the total capital) can actually increase the total capital available when combined with other {\sl losing} games. This shows that, for that particular class of games, redistribution of the capital is beneficial for everybody. The same conclusion arises when the redistribution goes from the richer players to the poorer. Finally, we would like to point out that ensemble of coupled Brownian motors have been considered in the literature\cite{rkbh99} and it would be interesting to see the relation they might have with the Parrondo type paradox described in this paper.

\noindent{\bf Acknowledgments:} This work is supported by the Ministerio de
Ciencia y Tecnolog{\'\i}a (Spain) and FEDER, projects BFM2001-0341-C02-01
and BFM2000-1108.


\begin{thebibliography}{000}
\bibitem{alternation} The same conclusion holds if games are played in some regular pattern such as AABBAABBAABB$\dots$, although for simplicity we will only consider the case of random alternation in this paper.
\end{thebibliography}
\end{document}